\documentclass[10pt, twocolumn,superscriptaddress,showpacs,preprintnumbers]{revtex4}
\usepackage{epsf} 
\usepackage{amsmath}
\usepackage{amssymb}
\usepackage{epsfig}
\usepackage{latexsym}
\usepackage{amsfonts}
\usepackage{dcolumn}
\usepackage{varioref}
\usepackage{ifthen}
\usepackage{graphicx}
\usepackage{amssymb,amsmath,amsthm}
\begin{document}
\parindent 0mm 
\setlength{\parskip}{\baselineskip} 
\thispagestyle{empty}
\pagenumbering{arabic} 
\setcounter{page}{1}
\mbox{ }
\preprint{UCT-TP-284/11}
%
\title	{QCD sum rule determination of the charm-quark mass}
\author{S. Bodenstein}
\affiliation{Centre for Theoretical \& Mathematical Physics, University of
Cape Town, Rondebosch 7700, South Africa}
\author{J. Bordes}
\affiliation{Departamento de F\'{\i}sica Te\'{o}rica,
Universitat de Valencia, and Instituto de F\'{\i}sica Corpuscular, Centro
Mixto Universitat de Valencia-CSIC}
\author{C. A. Dominguez}
\affiliation{Centre for Theoretical \& Mathematical Physics, University of
Cape Town, Rondebosch 7700, South Africa}
\affiliation{Department of Physics, Stellenbosch University, Stellenbosch 7600, South Africa}
\author{J. Pe\~{n}arrocha}
\affiliation{Departamento de F\'{\i}sica Te\'{o}rica,
Universitat de Valencia, and Instituto de F\'{\i}sica Corpuscular, Centro
Mixto Universitat de Valencia-CSIC}
\author{K. Schilcher}
\affiliation{Institut f\"{u}r Physik, Johannes Gutenberg-Universit\"{a}t
Staudingerweg 7, D-55099 Mainz, Germany}

\date{\today}
\begin{abstract}
QCD sum rules involving mixed inverse moment integration kernels are used in order to determine the running charm-quark mass in the $\overline{MS}$ scheme. Both the high  and the low energy expansion of the vector current correlator are involved in this determination. The optimal integration kernel turns out to be of the form $p(s) = 1 - (s_0/s)^2$, where $s_0$ is the onset of perturbative QCD. This kernel enhances the contribution of the well known narrow resonances, and reduces the impact of  the data in the range $s \simeq 20 - 25 \; \mbox{GeV}^2$. This feature leads to a substantial reduction in the sensitivity of the results to changes in $s_0$, as well as to a much reduced impact of the experimental uncertainties in the higher resonance region. The value obtained for the charm-quark mass in the $\overline{MS}$ scheme at a scale of $3 \; \mbox{GeV}$ is $\overline{m}_c (3 \; \mbox{GeV}) = 987 \; \pm 9 \; \mbox{MeV}$, where the error includes all sources of uncertainties added in quadrature.\\
\end{abstract}
\maketitle
\noindent
Progress on the theoretical \cite{QCD1}-\cite{QCD12}, as well as on the experimental information \cite{EXP1}-\cite{EXP4} on the vector current correlator has allowed for a considerable improvement on the accuracy of QCD sum rule determinations of the charm-quark mass \cite{K}-\cite{hoang}. The analysis of \cite{K} is based on inverse (Hilbert) moment QCD sum rules, requiring QCD knowledge of the vector correlator in the low energy, as well as in the high energy region. In  \cite{charm} an alternative approach was used which involves only QCD information at short distances, together with (a) a simple integration kernel $p(s) = 1 - s/s_0$ (local constraint), and (b) Legendre-type polynomial kernels (global constraint). In this paper we describe an improved analysis based on the  use of direct as well as inverse moment kernels of the form $p(s) = 1 - (s_0/s)^n$, with $n \geq 1$. These kernels enhance considerably the impact of the well known narrow resonances, as compared with e.g. a simple kernel $p(s) = 1/s^2$ , or $p(s) = 1- s/s_0$. They also provide a welcome stronger suppression of the contribution of data in the range $s \simeq 20 - 25 \; \mbox{GeV}^2$. In comparison with simple inverse moments without pinching, this means that results are  less  sensitive to assumptions about the onset of perturbative QCD (PQCD), as well as to the treatment of the higher resonance data. For instance, changes in $s_0$ in the range $s_0 \simeq 15 - 23 \; \mbox{GeV}^2$ lead to a variation in $\overline{m}_c(3\; \mbox{GeV})$ of only $4 \; \mbox{MeV}$ (for $n=2$) as opposed to a variation of $14 \; \mbox{MeV}$ for $p(s) = 1/s^2$, as used in \cite{K}\\ 

We  consider the vector current correlator
\begin{eqnarray}
\Pi_{\mu\nu} (q^2) &=& i \int d^4x \; e^{iqx} \langle 0| T(V_\mu(x) \; V_\nu(0))|0\rangle \nonumber \\ [.3cm]
&=& (q_\mu\; q_\nu - q^2 g_{\mu\nu})\; \Pi(q^2)\;,
\end{eqnarray}
where $V_\mu(x) = \bar{c}(x) \gamma_\mu c(x)$. From the residue theorem in the complex s-plane ($- q^2 \equiv Q^2 \equiv s$) it follows
\begin{eqnarray}
\int_{0}^{s_0}
p(s)\, \frac{1}{\pi} Im \,\Pi(s)\,ds &=& - \frac{1}{2\pi i}
\oint_{C(|s_0|)}
p(s) \,\Pi(s) \,ds \nonumber \\ [.3cm]
 &+& \text{Res}[\Pi(s) \,p(s),s=0]\;,
\end{eqnarray}
where $p(s)$ is an integration kernel, and 
\begin{equation}
Im\;\Pi(s) = \frac{1}{12 \pi} \;R_c(s) \;,
\end{equation}
with $R_c(s)$ the standard R-ratio for charm production. The PQCD expansion of $\Pi(s)$ at short distances can be written as
\begin{equation}
\Pi(s)|_{PQCD} = e_c^2 \;\sum_{n=0} \left( \frac{\alpha_s(\mu^2)}{\pi}\right)^n \; \Pi^{(n)}(s) \;,
\end{equation}
where $e_c = 2/3$ is the charm-quark electric charge, and
\begin{equation}
\Pi^{(n)} (s) = \sum_{i=0} \left(\frac{\bar{m}_c^2}{s}\right)^i \; \Pi^{(n)}_i\;,
\end{equation}
and $\overline{m}_c \equiv \overline{m}_c(\mu)$  is the running charm-quark mass in the $\overline{MS}$-scheme.
Up to order $\cal{O}$ $[\alpha_s^2 (\bar{m}_c^2/s)^6]$ the function $\Pi(s)_{PQCD}$   has been calculated in \cite{QCD1}, and exact results for $\Pi_0^{(3)}$ and $\Pi_1^{(3)}$ have been found in \cite{QCD2}. The function $\Pi_2^{(3)}$ is known exactly up to a constant \cite{QCD3}. At five-loop order $\cal{O}$$(\alpha_s^4)$ the full logarithmic terms for $\Pi_0^{(4)}$ may be found in \cite{QCD5}, and for $\Pi_1^{(4)}$ in \cite{QCD6}. Since there is incomplete knowledge at this loop-order we shall use the available information as a measure of the truncation error in PQCD. The low energy expansion of the vector correlator around $s=0$ can be written as
\begin{equation}
\Pi_{PQCD} (s) = \frac{3\, e_c^2}{16\, \pi^2}\; \sum_{n \geq 0} \overline{C}_n \; z^n\;,
\end{equation}
where $z = s/(4 \overline{m}_c^2)$. The coefficients $\overline{C}_n$ can be expanded in powers of $\alpha_s(\mu)$
\begin{eqnarray}
\bar{C}_n &=& \bar{C}_{n}^{(0)}+\frac{\alpha_{s}(\mu)}{\pi}\left(\bar{C}_{n}^{(10)}+\bar{C}_{n}^{(11)} l_{m}\right)
\nonumber \\ [.3cm]
&+&\left(\frac{\alpha_{s}(\mu)}{\pi}\right)^2 \left(\bar{C}_{n}^{(20)}+\bar{C}_{n}^{(21)} l_{m}+\bar{C}_{n}^{(22)} l_{m}^{2}\right) \nonumber \\ [.3cm]
&+& \left(\frac{\alpha_{s}(\mu)}{\pi}\right)^3 \left(\bar{C}_{n}^{(30)}+\bar{C}_{n}^{(31)} l_{m}+\bar{C}_{n}^{(32)} l_{m}^{2}\right.\nonumber \\ [.3cm]
&+& \left. \bar{C}_{n}^{(33)} l_{m}^{3}\right)+ \ldots
\end{eqnarray}
where $l_m\equiv \ln(\bar{m}_{c}^{2}(\mu)/\mu^2)$. Up to three loop level the coefficients up to $n=30$ of $\bar{C}_n$  are known \cite{QCD8}-\cite{QCD9}. At four-loop level we have $\bar{C}_0$ and $\bar{C}_1$ from \cite{QCD8}, \cite{QCD10}, $\bar{C}_2$ from \cite{QCD9}, and  $\bar{C}_3$ from \cite{QCD11}. We will choose $p(s)$ so that no coefficients $\bar{C}_4$ and above contribute to the residue at $s=0$, $\text{Res}[\Pi(s) \,p(s),s=0]$.

\begin{figure}
[ht]
\begin{center}
\includegraphics[height=2.5in, width=3.4in]{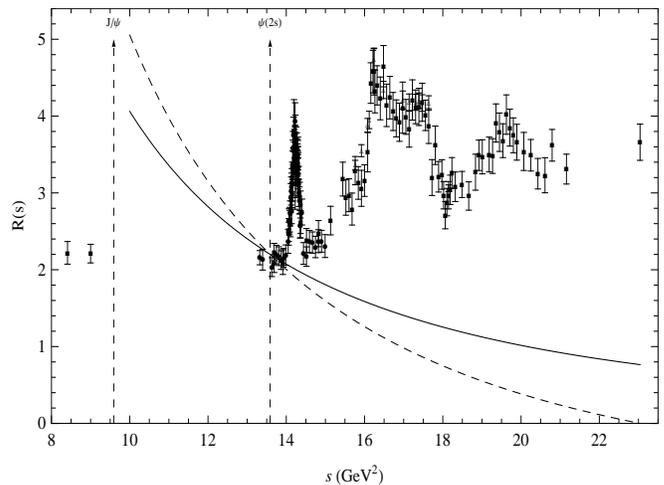}
\caption{Experimental data for the total  $R(s)$ ratio together with the optimal integration kernel, Eq.(8), with $n=2$ (dash curve), and $p(s) = 1/s^2$ (solid curve) normalized to coincide with the former at the position of the $\psi(2S)$ peak.}
\end{center}
\end{figure}
Apart from the quark mass, the fundamental QCD parameters are the running strong coupling $\alpha_s(\mu^2)$, and the gluon condensate. For the strong coupling we use the world average from \cite{BETH2}, which agrees with lattice QCD results \cite{LATT},  $\alpha_s(M_Z^2)= 0.1184 \pm 0.0007$. However, we will consider other values when comparing results for $\overline{m}_c$ with other analyses. 
In the non-perturbative sector the leading power correction in the Operator Product Expansion (OPE) involves the gluon condensate, i.e.
$\left<(\alpha_s/\pi)G^2\right>$ whose value has been extracted \cite{COND} from the ALEPH data on $\tau$-decays. While the gluon condensate is renormalization group invariant, its determination from QCD sum rules involves a difference between integrals of PQCD and integrated experimental data. This leads to an unavoidable dependence of the gluon condensate on the value of $\alpha_s$ used in the PQCD expression of the correlator. 
Extrapolating the  results of \cite{COND} to include current values of $\alpha_s$  \cite{BETH2}, \cite{PICH}, leads to
$\left<(\alpha_s/\pi)G^2\right>=(0.01 \pm 0.01)\;\text{GeV}^4$.
This large uncertainty in the value of the gluon condensate has only a very small impact on our results for $\bar{m}_c$.\\ 
Turning to the experimental data, we  follow closely the analysis of \cite{K}. For the first two narrow resonances we use the latest data from the Particle Data Group \cite{PDG}, $M_{J/\psi}= 3.096916 (11)\; \mbox{GeV}$, $\Gamma_{J/\psi \rightarrow e^{+} e^{-}} = 5.55 (14) \;\mbox{keV}$, $M_{\psi(2s)}= 3.68609 (4)\; \mbox{GeV}$, $\Gamma_{\psi(2s) \rightarrow e^{+} e^{-}} = 2.35 (4) \;\mbox{keV}$. These  two narrow resonances are followed by the open charm region where the contribution from the light quark sector $R_{uds}$ needs to be subtracted  from the total R-ratio $R_{tot}$. We perform this subtraction as in \cite{MC2}. In the region $3.97\;\mbox{GeV} \leq \sqrt{s} \leq 4.26\;\mbox{GeV}$ we only use CLEO data \cite{EXP4} as they are the most precise. In connection with the three data sets from BES \cite{EXP1}-\cite{EXP3}, we assume  that the systematic uncertainties are not fully independent and add them linearly, rather than in quadrature. However, we treat these data as independent from the CLEO data set \cite{EXP4}, and thus add errors in quadrature. There is no data in the region $s = 25 - 49 \;\mbox{GeV}^2$, and   beyond  there is CLEO data up to $s\simeq 110 \;\mbox{GeV}^2$. The latter data is fully compatible with PQCD.\\
We discuss next the integration kernels $p(s)$ in Eq.(2), which we choose as
\begin{equation}
p(s) = 1 - \left(\frac{s_0}{s}\right)^n \;,
\end{equation}

with $n \geq 1$. As discussed in \cite{K}, inverse moments  $p(s) = 1/s^n$ should not involve too large values of $n$. In fact, the convergence of PQCD deteriorates with increasing $n$, the gluon condensate contribution increases sharply for $n > 2$,  and the uncertainties in $\alpha_s$ and the renormalization scale $\mu$ have a greater impact on the total error of the charm-quark mass. On the other hand, direct kernels of the form $p(s) = s^n$, with $n \geq 1$, pose problems. Indeed, the high energy expansion of the vector correlator is incompletely known at $\mathcal{O}[\alpha^{3}_{s}]$, so that the greater the value of $n$ the greater is the contribution of the higher order mass corrections at this order in PQCD. Already with $n=1$ one would need a Pade approximation for the term $\mathcal{O}[\alpha^{3}_{s}\bar{m}_{c}^{6}]$. Hence, in order to avoid any approximation up to this order in PQCD we restrict ourselves to direct moments with $n=0$, and include inverse powers in an attempt to enhance the contribution of the well known narrow resonances, $J/\psi$ and $\psi(2S)$, and at the same time suppressing the broad resonance region.
We found that Eq.(8) with $n=2$ is the optimal kernel as explained next. In Fig.1 we show the experimental data for the ratio  $R(s)$ together with the kernel Eq.(8) with $n=2$ and $s_0 \simeq \; 23 \; \mbox{GeV}^2$, and the simple kernel $p(s)= 1/s^2$ normalized such that both kernels coincide at the peak of the second narrow resonance $\psi(2S)$, i.e. $s \simeq 13.6\; \mbox{GeV}^2$. One can easily appreciate that in comparison with the latter, the former kernel leads to a welcome enhancement of the weight of the $J/\psi$, as well as to a strong suppression of the broad resonance region, and particularly the region near the onset of the continuum. Quantitatively, the ratio of the area under the hadronic spectral function weighted with $p(s)$, in the narrow resonance region, ${\cal{I}}_{nr}$, and in the broad resonance region and beyond, ${\cal{I}}_{br}$, is ${\cal{I}}_{nr}/{\cal{I}}_{br} = 3.6$ for $p(s)=1/s^2$, and ${\cal{I}}_{nr}/{\cal{I}}_{br} = 7.7$ for Eq.(8) with $n=2$. Other values of $n$ lead to slightly less enhancement. In addition, the kernel Eq.(8) with $n=2$ leads to final results for $\overline{m}_c$ which are fairly insensitive to  the choice of $s_0$. For instance, in the range $s_0 \simeq \; 15.0 - 23.0 \; \mbox{GeV}^2$, $\overline{m}_c(3\; \mbox{GeV})$ changes by about $4.0 \;\mbox{MeV}$, while using the kernel $p(s)=1/s^2$ it changes by $14.0 \;\mbox{MeV}$.\\

\squeezetable
\begin{table}
\begin{ruledtabular}
\begin{tabular}{ccccc}
\multicolumn{4}{r}{$\overline{m}_c(3\; \mbox{GeV})$ (in MeV)} \\
\cline{2-5}
\noalign{\smallskip}
 Kernel  & $\bar{m}_{c}^{(0)}$ &$\bar{m}_{c}^{(1)}$  & $\bar{m}_{c}^{(2)}$ &$\bar{m}_{c}^{(3)}$   \\
\hline
\noalign{\smallskip}
$s^{-2}$ & 1129 & 1021 & 998 & 995  \\
$1-(s_0/s)^2$ & 1146 & 1019 & 991 & 987   \\
\end{tabular}
\caption{\footnotesize{Results for the charm-quark mass at different orders in PQCD, and for two integration kernels. The results for $p(s) = 1/s^2$ are obtained using slightly different values of the QCD parameters, and a different integration procedure as in \cite{K}.}}
\end{ruledtabular}
\end{table}


\begin{table}
\begin{ruledtabular}
\begin{tabular}{cccccccc}
\multicolumn{7}{r}{Uncertainties (in MeV)} \\
\cline{3-8}
\noalign{\smallskip}
 Kernel & $\bar{m}_c(3\,\mbox{GeV})$ & EXP & $\Delta \alpha_s$  & $\Delta \mu$ & NP & $s_0$ &  Total               \\
\hline
\noalign{\smallskip}
$s^{-2}$  &  995 \quad & \quad 9 \quad &\quad  3 \quad &\quad  1 \quad  &\quad  1 \quad &\quad 14 &\quad 17\\
$1-(s_0/s)^2$  &  987 \quad & \quad 7 \quad &\quad  4 \quad &\quad  1 \quad  &\quad  1 \quad &\quad 4 &\quad 9\\
\end{tabular}
\caption{\footnotesize{The various uncertainties due to the data (EXP), the value of $\alpha_s$ ($\Delta \alpha_s$), changes of $\pm 35 \%$ in the renormalization scale around $\mu= 3 \; \mbox{GeV}$ ($\Delta \mu$), the value of the gluon condensate (NP), and due to variations in $s_0$ ($s_0$).}}
\end{ruledtabular}
\end{table}

\begin{figure}
[ht]
\begin{center}
\includegraphics[height=2.5in, width=3.4in]{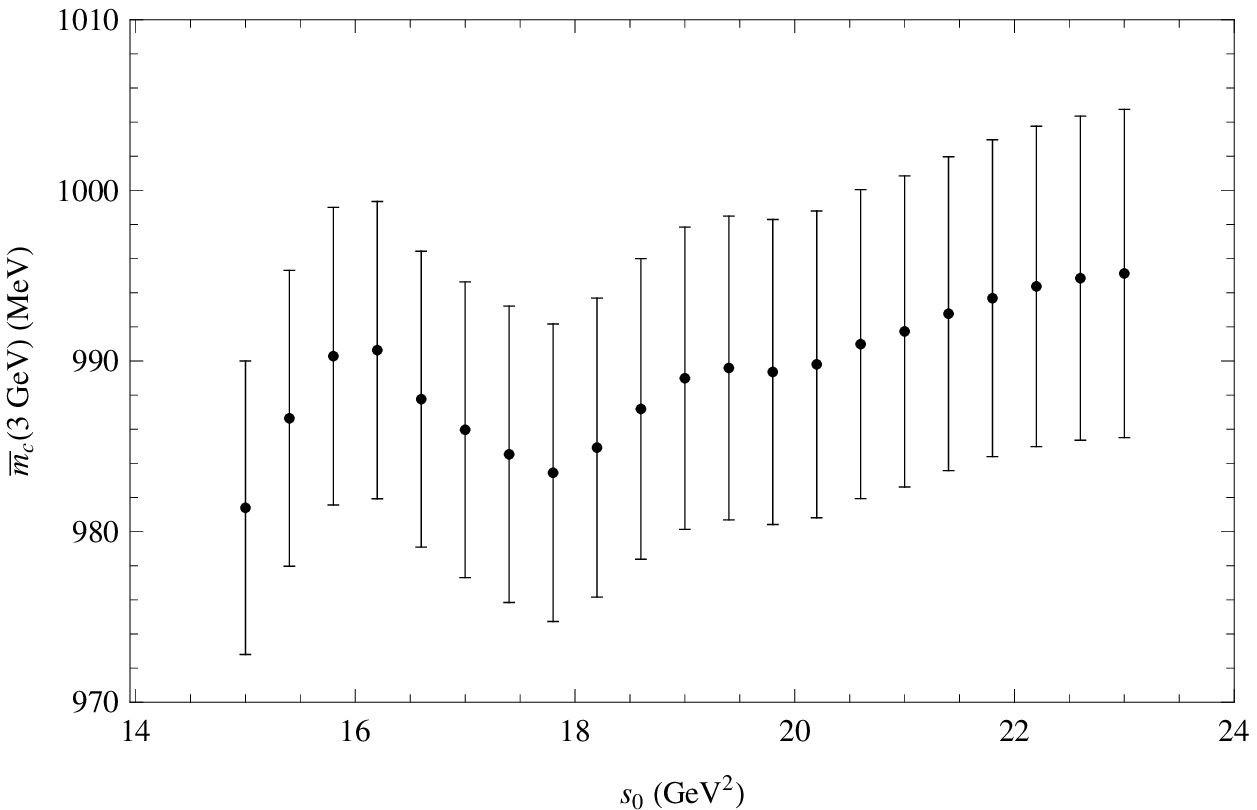}
\caption{Results for $\overline{m}_c(3\; \mbox{GeV})$ as a function of $s_0$ for the kernel $p(s) = 1/s^2$. The variation of $\overline{m}_c(3\; \mbox{GeV})$ in this range is up to $14 \; \mbox{MeV}$.}
\end{center}
\end{figure}

\begin{figure}
[ht]
\begin{center}
\includegraphics[height=2.5in, width=3.4in]{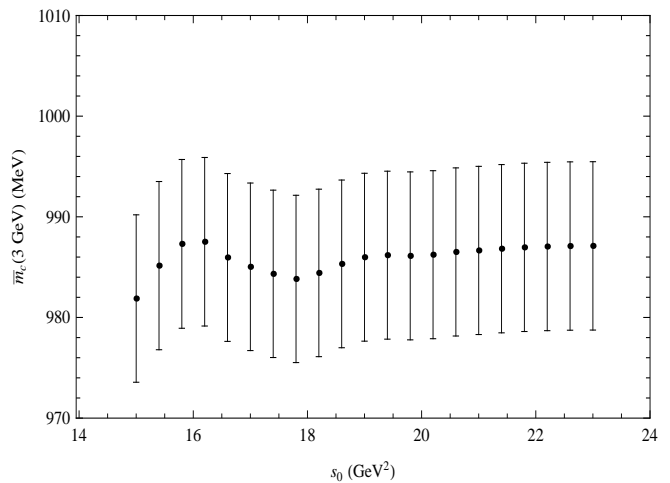}
\caption{Results for $\overline{m}_c(3\; \mbox{GeV})$ as a function of $s_0$ for the kernel Eq.(8) with $n=2$. The variation of $\overline{m}_c(3\; \mbox{GeV})$ in this range is up to $4 \; \mbox{MeV}$.}
\end{center}
\end{figure}

Proceeding to our determination we list in Table 1 the results for 
$\overline{m}_c(3\; \mbox{GeV})$ at different orders in perturbation theory, and using two integration kernels. The results for the kernel $p(s) = 1/s^2$ differ from \cite{K} as we use now a slightly different value of the strong coupling, of the gluon condensate, and of the $\psi(2S)$ parameters as given above, and we include the CLEO data \cite{EXP4}. The various errors associated to the final value of $\overline{m}_c(3\; \mbox{GeV})$ are given in Table 2. Results from Fixed Order Perturbation Theory (FOPT) are essentially the same as using Contour Improved Perturbation Theory (CIPT) to integrate around the circle of radius $s_0$ in the complex s-plane. In Fig.2 we show the results for $\overline{m}_c(3\; \mbox{GeV})$ as a function of $s_0$ for the kernel $p(s) = 1/s^2$, and in Fig. 3 for the kernel Eq.(8) with $n=2$, the latter exhibiting improved stability.

The convergence pattern in $\alpha_s$ of the PQCD integral as a function of $\overline{m}_c$ can be studied by computing 
\begin{equation}
{\cal{I}}(s_0) = - \frac{1}{2\pi i}
\oint_{C(|s_0|)}
p(s) \,\Pi(s) \,ds 
 + \text{Res}[\Pi(s) \,p(s),s=0] \;,
\end{equation}
with these integrals being functions of both $\overline{m}_c$ and $\alpha_s$. Using $\overline{m}_c(3\; \mbox{GeV}) = 987 \; \mbox{MeV}$, and Eq.(8) with $n=2$ we find reasonable convergence, i.e. ${\cal{I}}^{(0)} = 91.4 \;\mbox{GeV}^2$, ${\cal{I}}^{(1)} = 62.0 \;\mbox{GeV}^2$, ${\cal{I}}^{(2)} = 57.0 \;\mbox{GeV}^2$, and ${\cal{I}}^{(3)} =  56.3 \;\mbox{GeV}^2$, where the upper index in ${\cal{I}}^{(j)}$ indicates the power of $\alpha_s$.\\

Our final result using the optimal kernel, Eq.(8), with $n=2$ is
\begin{equation}
\bar{m}_c(3\, \mbox{GeV}) = 987\, \pm \, 9 \; \mbox{MeV} \;,
\end{equation}
in good agreement within errors with the result from inverse moment QCD sum rules \cite{K}, other recent determinations \cite{charm}-\cite{hoang}, \cite{MC2}, \cite{SIGNER}, as well as lattice QCD \cite{LATT}. Translated into a scale invariant mass, the above result gives $\bar{m}_c(\bar{m}_c) = 1278 \, \pm \,9  \; \mbox{MeV}$ for the value used here for the strong coupling.

This work was supported in part by the European FEDER and Spanish MICINN under grant MICINN/FPA 2008-02878, by the Generalitat Valenciana under grant GVPROMETEO 2010-056,  by DFG (Germany), and by NRF (South Africa).
One of us (KS) wishes to thank A. H. Hoang for helpful discussions.


\begin{thebibliography}{99}

\bibitem{QCD1}K. G. Chetyrkin, R. Harlander, J. H. K\"{u}hn, and M. Steinhauser,  Nucl. Phys. B {\bf 503}, 339 (1997).

\bibitem{QCD2}P. A. Baikov, K. G. Chetyrkin, and J. H. K\"{u}hn, Nucl. Phys. B (Proc. Suppl.) {\bf 189}, 49 (2009).

\bibitem{QCD3}K. G. Chetyrkin, R. Harlander, J. H. K\"{u}hn, Nucl. Phys. B {\bf 586}, 56 (2000).

\bibitem{QCD4}Y. Kiyo, A. Maier, P. Maierh\"{o}fer, and P. Marquard, Nucl. Phys. B {\bf 823}, 269 (2009). 

\bibitem{QCD5} P. A. Baikov, K. G. Chetyrkin, and J. H. K\"{u}hn, Phys. Rev. Lett. {\bf 101}, 012002 (2008).

\bibitem{QCD6} P. A. Baikov, K. G. Chetyrkin, and J. H. K\"{u}hn, Nucl. Phys. B (Proc. Suppl.) {\bf 135}, 243 (2004).

\bibitem{QCD7} K. G. Chetyrkin, J. H. K\"{u}hn, and M. Steinhauser, Phys. Lett. B {\bf 371}, 93 (1996); Nucl. Phys. B {\bf 482}, 213 (1996); ibid. B {\bf 505}, 40 (1997).

\bibitem{QCD8} R. Boughezal, M. Czakon, and T. Schutzmeier, Phys. Rev. D {\bf 74}, 074006 (2006); Nucl. Phys. B (Proc. Suppl.) {\bf 160}, 164 (2006).

\bibitem{QCD9}A. Maier, P. Maier\"{o}fer, and P. Marquard, Nucl. Phys. B {\bf 797}, 218 (2008); Phys. Lett. B {\bf 669}, 88 (2008).

\bibitem{QCD10}K. G. Chetyrkin, J. H. K\"{u}hn, and C. Sturm, Eur. Phys. J. C {\bf 48}, 107 (2006).

\bibitem{QCD11} A. Maier, P. Maierh\"{o}fer, P. Marquard, and A. V. Smirnov, Nucl. Phys. B {\bf 824}, 1 (2010). 

\bibitem{QCD12}A. H. Hoang, V. Mateu, and S. Mohammad Zebarjad, Nucl. Phys. B {\bf 813}, 349 (2009).

\bibitem{EXP1}J. Z. Bai et al. (BES 2000), Phys. Rev. Lett. {\bf 84}, 594 (2000).

\bibitem{EXP2}J. Z. Bai et al. (BES 2002), Phys. Rev. Lett. {\bf 88}, 101802 (2002).

\bibitem{EXP3} J. Z. Bai et al. (BES 2006), Phys. Rev. Lett. {\bf 97}, 262001 (2006).

\bibitem{EXP4} D. Cronin-Hennessy et al. (CLEO 2009), Phys. Rev. D {\bf 80}, 072001 (2009).

\bibitem{K} J. H. K\"{u}hn, M. Steinhauser, and C. Sturm, Nucl. Phys. B {\bf 778}, 192 (2007); K. G. Chetyrkin et al., Phys. Rev. D {\bf 80}, 074010 (2009).

\bibitem{charm} S. Bodenstein, J. Bordes, C. A. Dominguez, J. Pe\~{n}arrocha, and K. Schilcher, Phys. Rev. D {\bf 82}, 114013 (2010).

\bibitem{hoang} B. Dehnadi, A. H. Hoang, V. Mateu, S. Mohammad Zebarjad, arXiv: 1102.2264 (2011).

\bibitem{BETH2} S. Bethke, Eur. Phys. J. C {\bf 64}, 689 (2009).

\bibitem{LATT} C. T. H. Davies et al., HPQCD Collab., Phys. Rev. D {\bf 78}, 114507 (2008); C. McNeile, PQCD Collab., Phys. Rev. D {\bf 82}, 034512 (2010).

\bibitem{COND} C. A. Dominguez, and K. Schilcher, J. High Energy Phys. {\bf 0701}, 093 (2007).

\bibitem{PICH} A. Pich, arXiv:1101.2107.

\bibitem{PDG} K. Nakamura et al., Particle Data Group, J. Phys. G {\bf 37}, 075021 (2010).

\bibitem{MC2} A. Hoang, and M. Jamin,  Phys. Lett. B {\bf 594}, 127 (2004).


\bibitem{SIGNER} A. Signer, Phys. Lett.  B \textbf{672}, 333 (2009).
\end{thebibliography}
\end{document}